\documentclass[fleqn,usenatbib]{mnras}
\pdfoutput=1 
\usepackage{newtxtext,newtxmath}

\usepackage[T1]{fontenc}

\usepackage{graphicx}	
\usepackage{color}
\usepackage{cancel} 

\newcommand{\hmpc}{h^{-1}{\rm Mpc}}

\newcommand{\hgpc}{h^{-1}{\rm Gpc}}
\newcommand{\mbx}{\boldsymbol x}

\newcommand{\cdf}{{\rm CDF}}

\newcommand{\nn}{{\rm NN}}

\newcommand{\eq}[2]{\begin{align} \label{eq:#1} #2 \end{align}}

\newcommand{\dd}{{\rm d}}
\newcommand{\quijote}{\textsc{Quijote}\,}

\newcommand{\bq}{\boldsymbol q}
\newcommand{\bx}{\boldsymbol x}
\newcommand{\bk}{\boldsymbol{k}}
\newcommand{\bPsi}{\boldsymbol{\Psi}}

\title[Tracer-Field Cross-correlations with kNN CDFs]{Tracer-Field Cross-Correlations with $k$-Nearest Neighbor Distributions}

\author[Banerjee et. al]{
Arka Banerjee $^{1}$\thanks{E-mail: {\tt arka@iiserpune.ac.in}}, 
and Tom Abel $^{2,3,4}$\thanks{E-mail: {\tt tabel@stanford.edu}} \\
$^{1}$Department of Physics, Indian Institute of Science Education and Research,
Homi Bhabha Road, Pashan, Pune 411008, India\\
$^{2}$Kavli Institute for Particle Astrophysics and Cosmology, Stanford University, 452 Lomita Mall, Stanford, CA 94305, USA \\
$^{3}$Department of Physics, Stanford University, 382 Via Pueblo Mall, Stanford, CA 94305, USA \\
$^{4}$SLAC National Accelerator Laboratory, 2575 Sand Hill Road, Menlo Park, CA  94025, USA
}

\date{Accepted XXX. Received YYY; in original form ZZZ}

\pubyear{2022}
\begin{document}
\label{firstpage}
\pagerange{\pageref{firstpage}--\pageref{lastpage}}
\maketitle

\begin{abstract}
In astronomy and cosmology significant effort is devoted to characterizing and understanding spatial cross-correlations between points — e.g galaxy positions, high energy neutrino arrival directions, X--ray and AGN sources,  and continuous fields — e.g. weak lensing  and  Cosmic Microwave Background maps. Recently, we introduced the k-nearest neighbor ($k$NN) formalism to better characterize the clustering of discrete (point) datasets. Here we extend it to the point -- field cross-correlations analysis. It combines kNN measurements of the point data set with measurements of the field smoothed at many scales. The resulting statistics are sensitive to \textit{all} orders in the joint clustering of the points and the field. We demonstrate that this approach, unlike the 2-pt cross-correlation, can measure the statistical dependence of two datasets even when there are no linear (Gaussian) correlations between them. We further demonstrate that this framework is far more effective than the two point function in detecting cross-correlations when the continuous field is contaminated by high levels of noise. For a particularly high level of noise, the cross-correlation between halos and the underlying matter field in a cosmological simulation, between $10\hmpc$ and $30\hmpc$, is detected at $>5\sigma$ significance using the technique presented here, when the two-point cross-correlation significance is $\sim 1\sigma$.

Finally, we show that $k$NN cross-correlations of halos and the matter field can be well modeled on quasi-linear scales using the Hybrid Effective Field Theory (HEFT) framework, with the same set of bias parameters as are used for 2-pt cross-correlations. The substantial improvement in the statistical power of detecting cross-correlations using this method makes it a promising tool for various cosmological applications.

\end{abstract}




\section{Introduction}
\label{sec:intro}

Spatial correlations in the fluctuations or clustering of different datasets --- i.e. cross-correlations --- are increasingly being used in various cosmological analyses. Cross-correlations can be useful in multiple ways - from breaking parameter degeneracies, to mitigating the effects of various survey systematics \cite[see e.g.][for an overview]{2013arXiv1309.5388R}. In certain cosmological applications, one is interested in spatial correlations of fluctuations of number counts of a set of points, usually galaxies or quasar positions, and the fluctuations of a continuous field. These points can, in general, be thought to represent some sort of sampling of an underlying (cosmological) field, and we will therefore use the term ``tracers" to refer to them. Point-field or tracer-field cross-correlations are used, for example, in studies of galaxy-galaxy lensing, where galaxy positions are correlated with the shear field \cite[e.g.][]{2004AJ....127.2544S,2006MNRAS.368..715M,2006MNRAS.372..758M,2016PhRvD..94f3533P,DESY3gglensing,DESY3DMASS}. In combination with galaxy clustering, galaxy-galaxy lensing measurements have been crucial to pin down the (usually unknown) large scale bias of the galaxy sample under consideration, and consequently constrain the cosmological parameters of interest. Galaxy-galaxy lensing measurements have also been utilized to test models of gravity on large scales \cite[e.g.][]{2018MNRAS.479.3422A,2020A&A...642A.158B}. Similarly, galaxy positions have been cross-correlated with gravitational lensing of the Cosmic Microwave Background (CMB) \cite[see e.g.][and references therein]{2019PhRvD.100d3501O,2019PhRvD.100d3517O,2022arXiv220312440C}. Other examples include cross-correlations of the $21$cm emission signal with a sample of galaxies to isolate the cosmological signal in the presence of very large astrophysical foregrounds \citep{2022arXiv220201242C}. Given that foregrounds are expected to dominate the $21$cm auto-correlation signal, this particular cross-correlation will be crucial to exploit the potential  of the $21$cm surveys in terms of cosmology. Cross-correlation techniques have been applied to the study of detected gamma ray sources and weak lensing measurements \citep{2020PhRvL.124j1102A}, and galaxy catalogs with thermal Sunyaev-Zel'dovich (SZ) maps \citep{2020PhRvD.101d3525P}. Theoretically, cross-correlations between discrete tracers such as halos and voids with the continuous matter density field have been shown to be sensitive to certain unique effects in massive neutrino cosmologies, in both real and redshift space \citep{2014JCAP...03..011V, 2016JCAP...11..015B, 2018ApJ...861...53V,2020JCAP...06..032B,2022PhRvD.105l3510B}. Techniques to better characterize tracer-field cross-correlations can, therefore, be helpful for multiple areas of active research in cosmology.

In all cases highlighted above, the cross-correlations in the two datasets ultimately originate from the primordial fluctuations laid down during inflation \cite[see e.g.][]{1992PhR...215..203M}. These primordial fluctuations are consistent with being adiabatic and Gaussian with a nearly scale invariant spectrum \cite[see, e.g., ][and references therein]{2014A&A...571A..22P,2016A&A...594A..20P,2020A&A...641A..10P}, though searching for small departures from Gaussianity is an active area of research. At high redshifts, and on large (linear) scales even at low redshift, therefore, it is natural to consider only the Gaussian, or two-point, cross-correlations between two LSS datasets, as this captures the full extent of the correlations in the two datasets. In fact, it is quite common in cosmology to use the terms ``cross-correlations'' and ``two-point cross-correlations'' interchangeably, even though they can be different in principle. At late times, and on intermediate to small scales ($\lesssim 40 \hmpc$), nonlinear gravitational evolution naturally generates higher order cross-correlations. Specifically, the matter field itself is non-Gaussian on small scales at late times, as are tracers of this field, such as halos and galaxies. This leads to the emergence of non-trivial higher order terms, such as three-point cross-correlations \citep{2005A&A...432..783S}, that cannot be expressed in terms of the Gaussian cross-correlations, . Therefore, to fully capture tracer-field cross-correlations on quasi-linear to small scales, and maximally extract the cosmological information in these cross-correlations, it is essential to go beyond two-point cross-correlation statistics which are insensitive to higher order terms. In the context of higher order tracer-field cross-correlations, there exist various studies in the literature, pertaining to different science cases \cite[see e.g.][]{2005A&A...432..783S,2015JCAP...01..009A,2017A&A...606A.128R}, but here we explore a completely new approach.

Recently, the $k$-nearest neighbor ($k$NN) formalism has been put forward as a new measure for the clustering of discrete tracers of LSS \citep{Banerjee_Abel}. The $k$NN Cumulative Distribution Functions (CDFs) are formally sensitive to volume integrals of all connected $N$-point functions of the underlying continuous field sampled by the tracers, and have greater statistical power compared to the traditional 2-point function analysis. At the same time, measurements of these $k$NN-CDFs, using a tree structure to speed up nearest neighbor searches, are computationally cheap and scale as $N_{\rm query}\log N_D$, where $N_{\rm query}$ is the number of query points and $N_D$ the number of data points, and a single measurement procedure is sufficient to capture the clustering statistics over an entire range of scales. The $k$NN-CDFs for Gaussian fields are known analytically, which is advantageous as compared to other heuristic statistics one could construct. The framework was extended to also characterize the cross-correlations between two discrete datasets \citep{Banerjee_Abel_cross}. Cross-correlations measured using the $k$NN framework are sensitive to all possible connected $N$-point functions between the two underlying fields sampled by the two sets of tracers, and not just the linear, or Gaussian, cross-correlation measured by the two point cross-correlations. \citet{kNN_HEFT} demonstrated that the $k$NN distributions of biased cosmological tracers, such as halos, can be well modeled on quasi-linear scales using techniques that have been developed already to model the 2PCF on these scales, and that the $k$NN-CDFs can improve constraints on cosmological parameters by up to a factor of $3$, compared to the 2PCF, from the clustering of such tracers. \cite{kNN_SDSS} showed the first application of the $k$NN method to a cosmological dataset.

Given the statistical power of the $k$NN framework in capturing the clustering information of discrete data, it is important to ask if a similar setup can be developed to describe tracer-field spatial cross-correlations. In this paper, therefore, we focus on the methods needed to combine $k$NN measurements from a set of tracers with measurements on a continuous field, to draw out the spatial correlations in their clustering. In Sec.~\ref{sec:formalism}, we develop the formalism and measurement methods for computing tracer-field cross-correlations using the nearest neighbor measurements. In Sec.~\ref{sec:examples}, we present illustrative examples to demonstrate the statistical power of the nearest-neighbor tracer-field cross-correlations contrasted with two-point cross-correlations. In Sec.~\ref{sec:HEFT}, we turn to the question of modeling these cross-correlations on quasi-linear scales, and show that Hybrid Effective Field Theory (HEFT) \citep{modichenwhite} provides an appropriate framework to address this. Finally, in Sec.~\ref{sec:conclusions} we summarize our main results, and discuss some interesting aspects of the study.

\section{Formalism and Computational Framework}
\label{sec:formalism}

In this section, we lay out the formalism and  measurement methods for computing cross-correlations between tracers and continuous fields using nearest neighbor measurements. We first consider the ``continuum limit'' of $k$NN-CDFs to build intuition about the particular properties of the continuous field that can most easily be combined with the $k$NN measurements of the tracers. We then lay out the computational steps needed to measure these quantities, and how they are combined to characterize  cross-correlations in the spatial fluctuations of the two datasets.

\subsection{``Continuum limit'' of \textit{k}NN-CDFs}
\label{sec:continuum_limit}

A set of $N$ tracers, representing a local Poisson process on an underlying field, and distributed over a volume $V$, has a well defined number density $\bar n = N/V$. As demonstrated in \cite{Banerjee_Abel}, the nearest neighbor distributions depend explicitly on $\bar n$. Let us now consider the behavior of the $k$NN-CDFs as the mean number density $\bar n$ of the tracers tends to infinity. For this exercise, it is useful to connect the value of the $k\nn$-$\cdf$s at a given scale~$r$ to the probability of finding $\geq k$ data points in spheres of radius~$r$ \citep{Banerjee_Abel}:
\eq{cdf_pk}{\cdf_{k\nn} (r) = \mathcal P_{\geq k}(r) \, .}
This implies, for example, that the value of the 1-NN CDF at $r$ is equal to the probability of finding at least $1$ data point within spheres of radius $r$, averaged over the full volume under consideration. If the tracers represent a local Poisson process on some underlying continuous field whose (dimensionless) fluctuations are represented by $\delta (\mbx)$, the probability of finding exactly $k$ data points in a sphere of radius $r$, centered at $\mbx$, is given by 
\eq{poisson_sampling}{\mathcal P_{k}(\mbx) = \frac{\left[\lambda(\mbx)\right]^k}{k!} e^{-\lambda(\mbx)}\, ,}
where 
\eq{lambda_definition}{\lambda(\mbx) = \bar n V\Big(1+\delta_{r}(\mbx)\Big) \, ,}
and $\delta_{r}(\mbx)$ is the field smoothed on radius $r$ using a spherical tophat smoothing function  at location $\mbx$ . We now consider the limit of $\bar n \rightarrow \infty$, keeping the field $\delta_r$ unchanged, i.e. we sample the same underlying field with higher mean number density of tracers. From Eq.~\ref{eq:lambda_definition}, this implies exploring the limit of $\lambda \gg 1$. Since the Poisson distribution for $k$ is expected to peak near $\lambda$ we use Stirling's approximation on $k!$ for large $k$. Using this approximation, it is well known that the distribution can be well-approximated by a Gaussian in this limit:
\eq{Poisson_gaussian}{\mathcal P_k(\mbx) \simeq \frac{1}{\sqrt{2\pi (1/\lambda(\mbx)) }} \exp\Bigg[-\frac{\delta^2}{2(1/\lambda(\mbx))}\Bigg]\, ,}
where $\delta = (k-\lambda)/\lambda$. In the limit $\lambda \rightarrow \infty$, the Gaussian in Eq.~\ref{eq:Poisson_gaussian} is peaked at $0$ and has a vanishingly small width, so $\mathcal P_k(\mbx) \rightarrow \delta^D\left(k-\lambda(\mbx)\right)$, where $\delta^D$ represents the Dirac delta function. This is to be interpreted in the following way: the probability of finding exactly $k$ tracers in a sphere of radius $r$ located at $\mbx$ is non-zero when $\delta_r(\mbx)$ lies in a vanishingly small interval around $\delta_r^*$ where $\bar n V (1+\delta_r^*) = k$, i.e. it converges to $\delta^D(\delta_r(\mbx) - \delta_r^*)$. For any given sphere, therefore, the value of $k$ gets mapped onto a specific value of the enclosed overdensity as the mean number density of tracers is continuously increased.

The overall probability $\mathcal P_k(r)$ is obtained by integrating Eq.~\ref{eq:poisson_sampling} over all possible locations for the center of the sphere. Since this depends only on $\delta_r(\mbx)$, we can convert the integral over space into an integral over the possible values of $\delta_r(\mbx)$, i.e. the Probability Distribution Function (PDF) of $\delta_r$, $\phi(\delta_r)$:
\eq{integrated_pk}{\mathcal P_k(r) = \int \frac{\left(\lambda(\delta_r)\right)^k}{k!} e^{-\lambda(\delta_r)}\phi (\delta_r)\dd\delta_r \, .}
In the $\bar n \rightarrow \infty$ limit, therefore, this expression reduces to 
\eq{integrated_Pk_limit}{\mathcal P_k(r) \sim \int \delta^D(\delta_r-\delta_r^*)\phi(\delta_r)\dd \delta_r \propto \phi(\delta_r^*)\,.}
Therefore, the volume averaged probability of finding $k$ data points in spheres of radius $r$ is simply given by the PDF of the underlying continuous field $\delta_r$, from which the data points are sampled, at some $\delta_r=\delta_r^*$. Similarly, the probability of finding more than $k$ data points in spheres of radius $r$ will be mapped onto the probability of getting $\delta_r>\delta_r^*$, given the underlying PDF $\phi(\delta_r)$:
\eq{prob_gk}{\mathcal P_{\geq k}(r) \propto \int_{\delta_r^*}^{\infty} \phi(\delta_r) \dd r \,  = 1 - {\rm CDF}(\delta_r^*).}

Conceptually therefore, the continuum version of the $k$NN measurements are simply density thresholded evaluations of the CDF of the continuous field from which the tracers are sampled. Changing the value of $k$ at fixed $r$  corresponds to a change in the threshold value $\delta_r^*$ -- higher $k$ corresponds to a higher $\delta_r^*$. For a fixed k, but varying $r$, $\delta_r^*$ would also change. From Eq.~\ref{eq:lambda_definition}, it is clear that for larger $r$, and therefore, larger $V$, one needs a lower $\delta_r^*$ threshold for a fixed value of $k$ (or $\lambda$). Note that just like the discrete $k$NN-CDFs were formally sensitive to integrals of all connected $N$-point functions of the underlying continuous field \cite[see e.g.][]{White1979,Szapudi1993,Banerjee_Abel}, the full PDF (or equivalently, the CDF) is also formally sensitive to these higher order correlation functions \cite[see e.g.][]{PhysRevD.90.103519,PhysRevD.98.023508,2020MNRAS.495.4006U} \footnote{In the PDF literature, this is often referred to in terms of the higher order cumulants, as the PDF is obtained via a inverse Laplace transform of the Cumulant Generating Function (CGF).}. This will ensure that the cross-correlation formalism we develop in the next section is also guaranteed to capture the joint clustering at all orders. It is also interesting to note the connection of the quantity in Eq.~\ref{eq:prob_gk} to the zeroth Minkowski functional of continuous fields \citep{1989ApJ...340..647R, 1998MNRAS.297..355S}. We will explore the connection of nearest neighbor measurements to the other Minkowski functionals in future work.

\subsection{Tracer-Field Cross-Correlations}
\label{sec:cross_corr}
\begin{figure*}
	\includegraphics[width=0.9\textwidth]{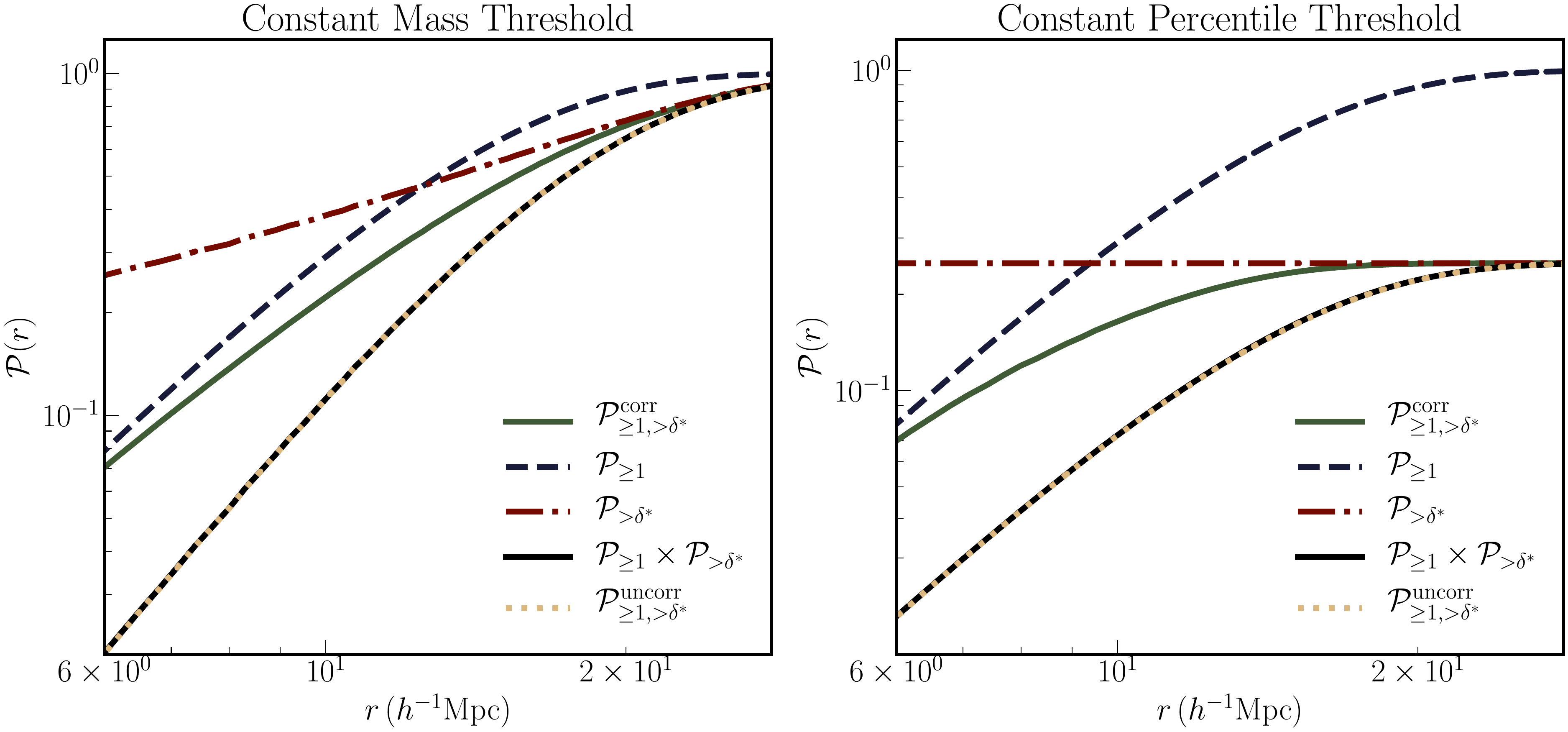}
	\caption{\textit{Left panel:} The solid green line represents the measurement of $\mathcal P_{\geq 1,>\delta_r^*}$, defined in Eq.~\ref{eq:joint_cdf_def}, where $\delta_r^*$ represents the threshold overdensity (see text for details), on a sample of $10^5$ halos and the underlying matter field from an $N$-body simulation at $z=0$, following the steps outlined in Sec~\ref{sec:calculations}. The dashed blue line represents $\mathcal P_{\geq 1}$ or the nearest neighbor CDF measured on the halos. The dash-dotted maroon line represents $\mathcal P_{>\delta_r^*}$, or the probability of the matter overdensity exceeding the threshold $\delta_r^*$ when smoothed on radius $r$, again calculated as outlined in the text. The density threshold, $\delta_r^*$, at radius $r$ is chosen such that $4/3 \pi r^3(1+\delta_r^*) = 4/3 \pi r_L^3$ with $r_L = 10 \hmpc$. The solid black line represents the $\mathcal P_{\geq 1}\times \mathcal P_{>\delta_r^*}$ - the expectation for the joint probability $\mathcal P_{\geq 1, >\delta_r^*}$ when the fluctuation in the tracer counts and the continuous field are statistically independent or uncorrelated. The dotted yellow line is obtained by measuring the joint probability $\mathcal P_{\geq 1, >\delta_r^*}$ on a set of $10^5$ halos from one realization and the matter field from a \textit{different} realization. As expected, the solid black line and the dotted yellow lines agree with each other. The departure of the solid green line from the solid black line, therefore, can be converted into a quantification of the cross-correlation between the halos and the matter field, as measured in the nearest neighbor framework. \textit{Right panel}: Same as the left panel, but with the threshold $\delta_r^*$ chosen to be the $75^{\rm th}$ percentile overdensity, as computed at each smoothing scale. This is reflected in the fact that $\mathcal P_{>\delta_r^*} = 0.25$, irrespective of the scale.}
	\label{fig:demo}
\end{figure*}

\cite{Banerjee_Abel_cross} showed that for two sets of discrete tracers, the spatial cross-correlations could be measured by considering the \textit{joint} $k$NN-CDFs -- $\cdf_{k_1,k_2}(r)$ -- which is equal to the (volume-averaged) probability of finding at least $k_1$ data points from the first set \textit{and} at least $k_2$ data points from the second set in spheres of radius $r$: $\mathcal P_{\geq k_1,\geq k_2}(r)$. Given the discussion in Sec.~\ref{sec:continuum_limit}, we would therefore expect that a measurement of the joint probability $\mathcal P_{\geq k, >\delta_r^*}(r)$ -- the probability of finding at least $k$ (tracer) data points \textit{and} the smoothed overdensity of the continuous field to cross threshold $\delta_r^*$ in spheres of radius $r$ -- to capture the cross-correlations between tracers and a continuous field. If the positions of the tracers represent a Poisson sampling of some (presumably different) continuous field $\tilde \delta (\mbx)$, then for any value of $k$, 
\eq{joint_prob_def}{\mathcal P_{k, >\delta_r^*} = \int_{\delta_r^*}^{\infty} \frac{\left(\lambda(\tilde \delta_r)\right)^k}{k!} e^{-\lambda(\tilde \delta_r)} \phi(\tilde \delta_r, \delta_r)\dd \tilde \delta_r \dd \delta_r \, ,}
where $\phi(\tilde \delta_r, \delta_r)$ is the full joint probability distribution of the two fields $\tilde \delta$ and $\delta$ when smoothed on scale $r$, and includes all information about how their fluctuations are correlated with each other. Therefore, just as in \cite{Banerjee_Abel_cross}, a measurement of $\mathcal P_{\geq k, >\delta_r^*}(r)$,
\eq{joint_cdf_def}{\mathcal P_{\geq k, > 
\delta_r^*} = \mathcal P_{>\delta_r^*} - \sum_{k^\prime<k}\mathcal P_{k^\prime, >\delta_r^*}}
with $\delta_r^*$ as the chosen threshold, will be sensitive not just to the linear, or Gaussian, correlations in the density fluctuations of the tracers and the continuous field, but to correlations in fluctuations at all orders. We will demonstrate this through an toy example in Sec.~\ref{sec:beyond_Gaussian}.

When the fields $\tilde \delta$ and $\delta$ are completely statistically independent, i.e. the joint distribution function can be factorized into to separate PDFs, $\phi(\tilde \delta_r, \delta_r) \propto \phi_1(\tilde \delta_r) \phi_2(\delta_r)$,
\eq{procuct}{\mathcal P_{\geq k, >\delta_r^*} = \mathcal P_{\geq k} \times \mathcal P_{>\delta_r^*} \, .}
Any departures from the above condition can be treated as a measure of the existence of correlations in the fluctuations of the two fields. We will typically use $\mathcal P_{\geq k, >\delta_r^*}/(\mathcal P_{\geq k} \times \mathcal P_{>\delta_r^*})$ as that measure, as it also connects trivially to the residuals we plot for many of the figures in the following sections. As the examples below demonstrate, ``positive" correlations between the tracers and the field are typified by this ratio being greater than $1$. We also note that we will typically not probe very deep into the tails of the distributions, and restrict ourselves to radial scales where the denominator in the formula is above is $\neq 0$.

\subsection{Computational setup}
\label{sec:calculations}

We now discuss the computational setup  for computing $\mathcal P_{\geq k, >\delta_r^*}$ given a set of discrete tracers and a continuous field\footnote{Since we will mainly be concerned with data from $N$-body simulations in this paper, we consider the case of periodic boundary conditions. However, the measurement techniques discussed below can be applied to situations with non-periodic boundary conditions with some minor modifications.}. 

\begin{enumerate}
	\item Define a regular $3$-dimensional grid with $N_g^3$ grid points. 
	\item Build a tree from the set of tracer positions. Using this tree, the distances to the $k$-nearest neighbor data points from each grid point can be determined in $\sim N_g^3\log N$ time. For each $k$ under consideration, the distances are sorted to produce the empirical $\cdf_{k\nn}(r)$. In the limit of large $N_g^3$, this approaches $\mathcal P_{\geq k}(r)$. Note that these measurements need to be made only once to produce the values of $\mathcal P_{\geq k}(r)$ over a range of $r$ \cite[see ][for more details]{Banerjee_Abel}.
	\item Smooth the continuous field on a smoothing scale $r$ using a $3$-dimensional spherical top-hat function. The smoothing is done in Fourier space. Note that while we call this a continuous field, in a simulation, it is typically also defined on a grid. This grid on which the field is defined may not be the same as the one defined in step (i). In case they are different, the values of the smoothed field are interpolated to the grid points defined in (i).
	\item For a given $k$, and a threshold $\delta_r^*$, compute the fraction of the $N_g^3$ grid points for which the the $k$-th nearest neighbor tracer point lies at a distance $<r$ \textit{and} the smoothed field, interpolated to the grid point, exceeds the threshold of $\delta_r^*$. In the limit of large $N_g^3$, this fraction approaches $\mathcal P_{\geq k, > \delta_r^*}$. 
	\item Compute the fraction of grid points for which the smoothed field, interpolated to the grid point, exceeds the threshold of $\delta_r^*$. In the limit of large $N_g^3$, this fraction approaches $\mathcal P_{> \delta_r^*}$. 
	\item Repeat steps (iii)-(v) for different values of the smoothing scale $r$.
\end{enumerate}

To isolate the part of the above measurements that come from cross-correlations, we will often focus on the quantity $\mathcal P_{\geq k, > \delta_r^*}/(\mathcal P_{\geq k} \times \mathcal P_{> \delta_r^*})$ in the later sections. When this quantity is $1$ (within appropriate error bars, which we will also discuss in the later sections), the spatial fluctuations in the tracer number density and the overdensity of the field are statistically independent. Any statistically detectable departure from $1$ is a measure of the correlation in the fluctuations. By measuring this quantity at different $r$, the cross-correlation can be probed as a function of scale.

To illustrate what these look like, we plot the results of some example measurements in Fig.~\ref{fig:demo}. For discrete tracers, we choose $10^5$ halos from the $z=0$ snapshot of an $N$-body simulation, from the fiducial cosmology runs of the \textsc{Quijote} simulations\footnote{https://quijote-simulations.readthedocs.io/en/latest/} \citep{2020ApJS..250....2V}, with a $(1\hgpc)^3$ volume\footnote{We will typically use this number density for tracers throughout the paper, since it is roughly that for cosmologically relevant tracer samples, like the BOSS CMASS sample}. For the continuous field, we use the matter field, as defined by all particles ($512^3$), from the same simulation and at the same redshift. The query points, from which the nearest-neighbor distances are measured, and around which we consider the value of the smoothed field, are placed on a $256^3$ grid spanning the simulation volume. We consider two different schemes for choosing the threshold value $\delta_r^*$. For the first scheme, shown in the left panel of Fig.~\ref{fig:demo}, at each $r$, we define $\delta_r^*$ such that $4/3\pi r^3(1+\delta_r^*) = 4/3 \pi r_L^3$, with $r_L = 10\hmpc$. Since the background density is assumed to be uniform, this is equivalent to defining $\delta_r^*$ in terms of a constant mass $M_0 = 4/3 \pi r_L^3\bar \rho$. For small values of $r$, the threshold $\delta_r^*$ is then a large positive value, and very few points cross the threshold ($\mathcal P_{>\delta_r^*}\sim 0$). As $r$ increases, $\delta_r^*$ gets smaller, crosses $0$ at some points, and then assumes negative values. At some point, all points cross the threshold ($\mathcal P_{>\delta_r^*} \sim 1$). This is shown by the dot-dashed maroon line. The dashed blue line represents the CDF of the first nearest neighbor distribution, or equivalently $\mathcal P_{\geq 1}(r)$. The solid green line represents the measurement of $\mathcal P_{\geq 1, >\delta_r^*}$. The solid black line represents the product of the (dot-dashed) maroon and (dashed) blue lines, and is the expectation for $\mathcal P_{\geq 1, >\delta_r^*}$ if the tracer and field fluctuations were statistically independent. The dotted yellow lines represent the measurement of $\mathcal P_{\geq 1, >\delta_r^*}$ for halos one one simulation, and the matter field from a \textit{different} simulation at the same cosmology, i.e. when the spatial fluctuations are indeed physically expected to be statistically independent. As expected, the (solid) black and (dotted) yellow lines agree with each other. The difference of the (solid) green line from the black line captures the cross-correlation of the halos and matter field in a single realization. 

The right panel of Fig.~\ref{fig:demo} represents the same set of measurements as in the left panel, but for a different scheme of choosing the overdensity threshold $\delta_r^*$. Instead of using a fixed $M_0$ to define the threshold, we use a fixed overdensity \textit{percentile} across all values of $r$. In the plotted example, we define $\delta_r^*=\delta_{r,75}$, the value of the $75^{\rm th}$ percentile of $\delta_r$. $\mathcal P_{>\delta_r^*} =0.25$ in this case, irrespective of the smoothing scale $r$, and this can be seen clearly from the (dot-dashed) maroon line. All other lines represent the set of measurements as in the left panel. This is to demonstrate that the cross-correlation can be captured through either choice, as long as they are applied consistently through the full analysis\footnote{One can also define the threshold in other ways -- using a fixed overdensity value, for example.}. The particular choice of the threshold scheme can depend on the application and the specific question of interest, and in Sec.~\ref{sec:examples} and Sec.~\ref{sec:HEFT}, we present examples with either of these choices.


\section{Example applications}
\label{sec:examples}

In this section, we present examples applications of the formalism from Sec.~\ref{sec:formalism}. First, using a toy example, we demonstrate that the formalism can be used to detect correlation between a set of tracers and a continuous field even in the case where the linear (Gaussian) correlations are set to $0$,  and cannot therefore, be detected by the two-point cross-correlations. Second, we demonstrate that in situations where the continuous field is contaminated by high noise levels, the nearest neighbor cross-correlation outperforms the two-point cross-correlation function in terms of signal-to-noise of the detection.

\begin{figure}
	\includegraphics[width=0.45\textwidth]{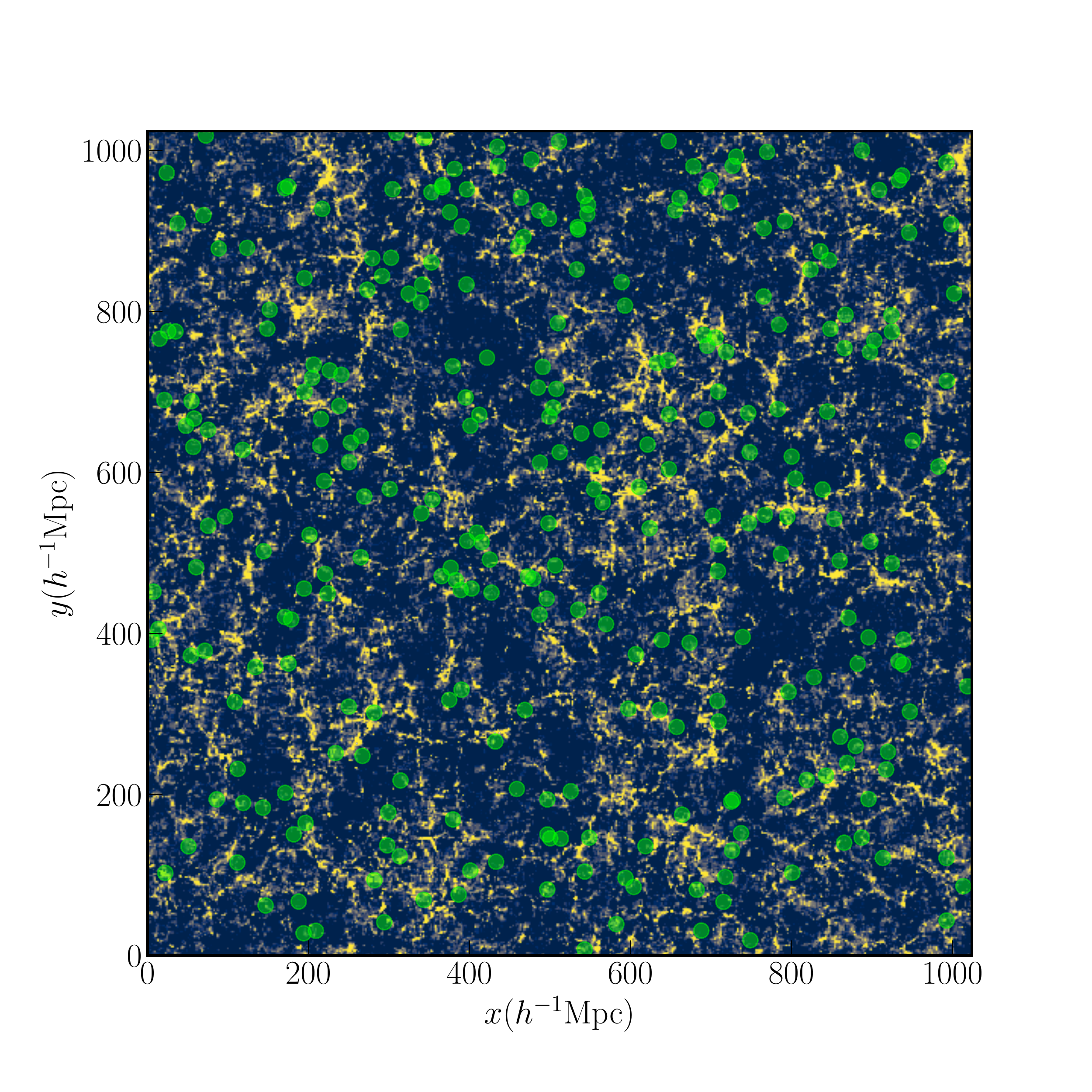}
	\caption{The green circles represent the positions of a random sub-selection of halos in a simulation, chosen from the $10^5$ most massive objects in a $(1\hgpc)^3$ volume that lie within a $100\hmpc$ projection along the $z$-axis. The underlying field is a $100\hmpc$ projection of the field $\delta_X$ from Sec.\ref{sec:beyond_Gaussian} constructed such that its two-point cross-correlation with the matter field from the simulation is $0$ (see text for details). The distribution of the green dots and the $\delta_X$ are not completely independent though, even visually, where the gaps in the green dots line up with the blue regions of the image. As demonstrated in Fig.\ref{fig:beyond_Gaussian}, the nearest neighbor cross-correlations are sensitive to this dependence, even though the 2-point function is not.}
	\label{fig:realizations}
\end{figure}

\subsection{Detecting beyond-Gaussian cross-correlations}
\label{sec:beyond_Gaussian}

\begin{figure*}
	\includegraphics[width=0.9\textwidth]{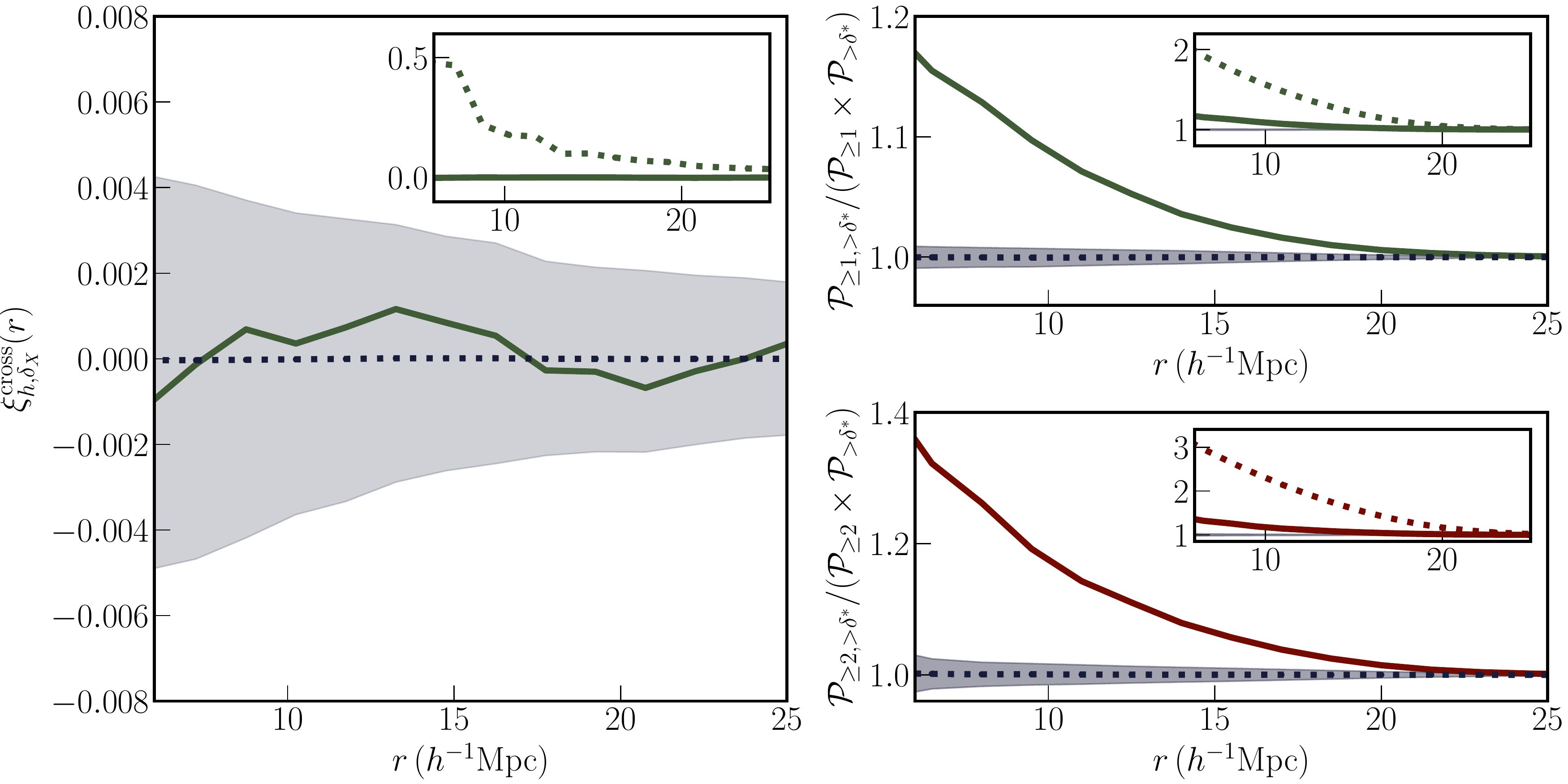}
	\caption{\textit{Left panel:}  The solid green line represents the measurement of the 2-point cross correlation function between the $10^5$ most massive halos in a simulation box of $(1\hgpc)^3$ with the field $\delta_X$ (defined in the text), constructed to have $0$ cross-$P(k)$ with the underlying matter density field $\delta$ in the simulation. The dotted blue line represents the mean (from $100$ different realizations) of the same measurement when halos from one realization is cross-correlated with the matter field from a different realization, i.e. they are completely uncorrelated. The shaded blue region represents the scatter in the measurement around the mean. The fact that the green line falls within the blue shaded region implies that the 2-point cross-correlation fails to detect cross-correlation between the halo positions and $\delta_X$ at any level of significance. For reference, the dashed line in the inset shows the size of the cross-correlation signal between the halos and the original matter field $\delta$, while the solid line is the same as in the main panel. \textit{Right panels}: Cross correlations between the halos and $\delta_X$, but now computed using the nearest neighbor method, with the top right panel showing the result for the $k=1$ nearest neighbor, and the bottom panel for the $k=2$ nearest neighbor. Once again, the blue dotted line and shaded regions represent the mean and the scatter of the measurement when the halos are cross-correlated (using the NN setup) with the matter field from a different realization. The $k$NN  measurements are clearly able to detect the correlations between the fluctuations in the halo number counts and the $\delta_X$ field even though the two-point cross-correlation is not detected. For reference, the dotted lines in the inset of each panel shows the size of the cross-correlation signal of the halos with the original matter field. While the total signal does reduce  (solid lines in each inset) when the two-point cross-correlation is removed, the total cross-correlation is still statistically significant with the NN method.}
	\label{fig:beyond_Gaussian}
\end{figure*}

Most applications in cosmology focus on two-point cross-correlations, or interchangeably, the cross power-spectrum, of the spatial fluctuations in two datasets. For Gaussian random fields, and their tracers, the two-point cross-correlation captures the correlation between the two datasets. However, even for fields that were initially well-approximated as Gaussian random fields, gravitational evolution in the quasi-linear and nonlinear regimes will generate non-trivial higher order correlations between the fields, and any tracers of these fields. The two-point cross-correlation, or cross $P(k)$, is insensitive to these higher order correlations, and continues to capture only the Gaussian part of the correlation. For realistic cosmological fields, though, there is still sufficient information in the Gaussian cross-correlation, even on the quasi-linear and nonlinear scales. The nearest neighbor cross-correlations between tracers and fields, on the other hand, are formally sensitive to cross-correlations at all orders.

To demonstrate this point, and contrast it with the two-point cross-correlation, we consider the following toy example. Once again, we consider the $10^5$ most massive halos from one of the fiducial \quijote simulations, along with the matter overdensity field $\delta(\mbx)$ grid at $z=0$. The Fourier transform of the overdensity field, $\tilde \delta (\bk)$, can be written as 
$\tilde \delta (\bk) = a(\bk) + i b (\bk)$ for each $\bk$. Next, we consider the field $\tilde \delta_X(\bk) = b(\bk) - i a(\bk)$, created by swapping the real and imaginary parts of $\tilde \delta (\bk)$ along with a sign change for the imaginary part\footnote{To ensure that $\delta_X(\mbx)$ itself is real, we need to ensure that $\tilde \delta_X(-\bk) = \tilde \delta_X^*(\bk)$. In our actual implementation, we use Fourier Transform routines for real arrays, where $\tilde \delta_X(\bk)$ is defined only for $k_z>0$, and the condition above is implicitly assumed for $k_z<0$ modes.}. Given the definition of the cross-power spectrum of two fields 
\eq{cross_pk}{P^{\rm cross}_{\delta_1, \delta_2}(k) \propto &\Big \langle \tilde \delta_1^*(\bk)\tilde \delta_2(\bk) \Big \rangle_{|\bk| = k}\nonumber \\  = &\Big \langle \Re\left(\delta_1(\bk)\right) \Re \left(\delta_2(\bk)\right) + \Im\left(\delta_1(\bk)\right)\Im\left(\delta_2(\bk)\right) \Big \rangle \,,}
it is clear that $P^{\rm cross}_{\delta, \delta_X}(k) =0$ for all $k$\footnote{Note, however that the imaginary part of the cross power spectrum is not $0$ here.}. Also note that given the way $\tilde\delta_X$ is constructed, $P^{\rm auto}_{\delta_X} (k)\simeq P^{\rm auto}_{\delta}(k)$, i.e. the two fields have similar levels of ``clustering''. Once we transform back to configuration space, we have a field $\delta_X(\mbx)$, which has no linear or Gaussian cross-correlation with the original field $\delta(\mbx)$. In practice, we define all quantities on a $512^3$ grid, and use a discrete FFT to move to Fourier space instead of the continuous Fourier transform, but this does not affect any of the arguments presented above. It is important to note, however, that the field $\delta_X(\mbx)$ should not be thought of as a physical overndensity field, as it is not guaranteed to be bounded below by $-1$. Therefore, such a field is unlikely to have a cosmological counterpart, and is only used here to demonstrate the theoretical sensitivity of the nearest neighbor measurements to higher order correlations. We show a view of a $100\hmpc$ projection (projected along the $z$-axis) of the $\delta_X$ field in Fig.~\ref{fig:realizations}. On the same figure, we plot the positions of a random sub-selection of halos --- from the $10^5$ most massive ones throughout the full volume --- that fall within this projection using green circles. Even though we have removed the linear cross-correlation between $\delta_X(\mbx)$ and $\delta(\mbx)$, and the halos are obviously highly correlated with the latter, it is clear by eye that the fluctuations of $\delta_X$ and the halo positions are not completely independent of each other. This is especially clear in the regions where $\delta_X$ has a low value (represented by the blue regions of the image), where there are also fewer halos.

To capture this quantitatively, we compute $\xi^{\rm cross}_{h, \delta_X}(r)$, the 2-point cross-correlation between the halo positions, and $\delta_X$. In practice, this is computed by averaging $\delta_X$ in spherical shells at radius $r$ and thickness $dr$, around the positions of all halos in the sample. We also compute $\xi^{\rm cross}_{h, \delta_i}$, where $\delta_i$ represents the $z=0$ matter overdensity field computed from each of $100$ other realizations at the same cosmology. Since the auto-clustering of the $\delta_X$ field is roughly at the same level, as that of the matter field at the same redshift and since we expect no cross-correlations between the positions of halos from one realization with the matter field from a different realization, we use these $100$ measurements to estimate the error bar on the measurement of $\xi$ in the presence of the discreteness noise of the halos. On the left panel of Fig.~\ref{fig:beyond_Gaussian}, the solid green line represents $\xi^{\rm cross}_{h, \delta_X}(r)$. The shaded grey region indicate the realization-to-realization scatter in the measurement of $\xi^{\rm cross}_{h, \delta_i}$ around the mean of the measurement from $100$ realizations(dotted blue lines). The fact that the green line falls within the grey shaded region across all scales indicates that there are no detectable linear  (Gaussian) cross-correlations between the positions of halos and the fluctuations of the field $\delta_X$. This is entirely expected, since we constructed $\delta_X$ to have no linear cross-correlations with the physical matter overdensity field $\delta$, and the positions of halos will, at least on quasi-linear scales, trace some biased version of $\delta$.

\begin{figure*}
	\includegraphics[width=0.9\textwidth]{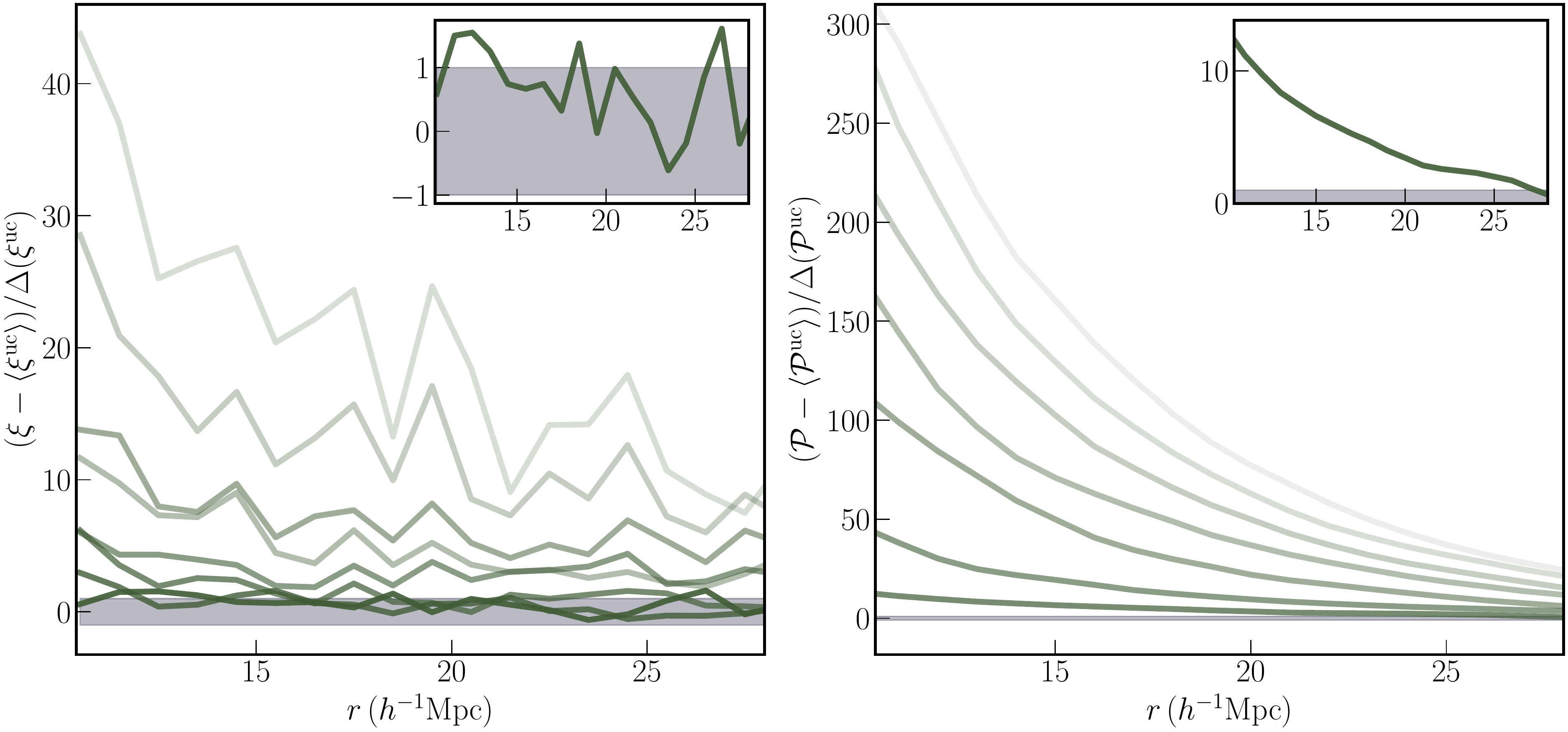}
	\caption{\textit{Left panel:}  Two-point cross-correlation measurements between halos and the underlying matter field in the presence of noise added to the matter field, parameterized by $f_{\rm noise}$ defined in Eq.~\ref{eq:noisy_field}. Darker lines correspond to higher $f_{\rm noise}$. We plot the difference between the measurement in the correlated case and the mean of $100$ uncorrelated cases (where the halos and the matter field are taken from different realizations), divided by the scatter in the measurements from the $100$ uncorrelated measurements at each level of $f_{\rm noise}$. The shaded grey region represents the $1\sigma$ scatter in the uncorrelated measurements -- these are horizontal lines across different $f_{\rm noise}$ given the quantity that is being plotted. The inset zooms in on the $f_{\rm noise}=0.95$ curve. \textit{Right panel}: Same as the right panel, except the halo-matter cross-correlations are measured via the nearest-nearest neighbor method, using $\mathcal P_{\geq 1, >\delta_r^*}$, where the threshold $\delta_r^*$ is chosen as the $50^{\rm th}$ percentile in the smoothed density at each radius $r$. As the inset demonstrates, there is a clear cross-correlation signal even for $f_{\rm noise}=0.95$ , in contrast to the two-point measurements in the right panel, where the measurement lies almost entirely within the shaded region. It should be noted that there are significant correlations between the measurements at various radii for the $k$NN cross-correlations, so the total signal-to-noise needs to be quantified accounting for these. This is presented in the text through the $\chi^2$ values for $f_{\rm noise}=0.95$ --- $20.58$ for the two-point cross-correlation, and $127.97$ for the $k$NN cross-correlation, with $19$ degress of freedom in each case.}
	\label{fig:noise}
\end{figure*}

On the right hand panel of Fig.~\ref{fig:beyond_Gaussian}, we plot the measurements of the cross-correlation on the same scales, but now calculated using the setup outlined in Sec.~\ref{sec:formalism}. The top right panel is for the $k=1$ nearest neighbor, while the bottom right panel is for $k=2$ nearest neighbor measurements. The query points are placed on a $256^3$ grid. Since the  field under consideration $\delta_X$ does not represent a physical density field, we use a percentile density threshold at each smoothing scale $r$ instead of a constant mass density threshold. As mentioned previously, we focus on the quantity $\mathcal P_{\geq k, > \delta_r^*}/(\mathcal P_{\geq k}(r) \times \mathcal P_{> \delta_r^*})$ which is a measure of the cross-correlation. This is plotted with the solid green line on the top right panel, and the solid maroon line in the bottom right panel. As with the two-point cross-correlation, we compute the nearest neighbor cross-correlations using halos from one realization as the set of tracers, and the density field from each of $100$ other realizations. The blue dotted line represents the mean of these measurements, while the shaded grey region represents the scatter around the mean measurement of this quantity from the $100$ realizations (dotted blue line). As expected, the dotted blue line is  consistent with $1$, meaning that the tracers from a given realization are, on average,  independent of the density fluctuations from a different realization. However, the green line is significantly different from $1$, even after accounting for the measurement errors represented by the grey bands. This implies that the nearest neighbor and smoothed field measurements are able to detect correlations in the fluctuations of the number counts of tracers and the in the $\delta_X$ field. Since the linear cross-correlations have been removed by construction (and as verified in the left panel), these cross-correlations can only be generated at higher orders.

For reference, we plot the cross-correlation signal between the halos and the actual matter field from the simulations, $\delta(\mbx)$ using dotted lines in the inset of each panel. The solid lines in each inset is the same as that in the main panels. The comparison between the solid and dotted lines serves as a rough estimate for the fraction of the cross-correlation signal lost when the two-point cross-correlation is removed by construction. For the two-point cross-correlation, of course, there is no remnant signal. For the nearest neighbor cross-correlations, a large fraction of the total cross-correlation is removed through the two-point, but that does not account for the entire signal - the remaining cross-correlation is still statistically significant.

Through this example, therefore, we have demonstrated that the cross-correlation formalism in Sec.~\ref{sec:formalism} is able to capture higher order correlations in the spatial fluctuations of number counts of tracers and the fluctuations of a continuous field --- something that the two-point cross-correlation (by design) is insensitive to. While the particular example presented here has no immediate cosmological applications, the sensitivity to higher order cross-correlations has many applications in cosmology --- from higher sensitivity to  cosmological parameters, to tests of various survey systematics.

\subsection{Detecting cross-correlations in the presence of noise}
\label{sec:noisy_fields}

In this section, we explore a more realistic cosmological application, which illustrates the greater statistical power of defining tracer-field cross-correlations through nearest neighbor measurements. We consider the problem of robustly detecting the spatial cross-correlation signal between a set of tracers, and an underlying field which is heavily contaminated with noise. This has potential applications, for example, in the detection of the clustering of the cosmological 21cm emission line signal, which is expected to be dominated by astronomical foregrounds, through its cross-correlation with a galaxy sample. \cite{2022arXiv220201242C} have already detected such a signal through a stacking measurement with eBOSS galaxies and quasars \citep{2016AJ....151...44D}, but techniques that  enhance the total signal-to-noise of the cross-correlation detection can help extract more cosmological information from the same underlying datasets.

We start with the matter overdensity field, $\delta_m$, from one realization of the \textsc{Quijote} suite at the fiducial cosmology, and at $z=0$. The field is defined on a $512^3$ grid. We define a new field 
\eq{noisy_field}{\tilde \delta = (1-f_{\rm noise}) 
\delta_m + f_{\rm noise}\delta_{\rm noise}\,,}
for each grid point, with $1+\delta_{\rm noise}$ drawn (independently) from a lognormal distribution with  $\mu = 0$ and $\sigma=3$\footnote{We have also tested with other distributions, such as a uniform distribution between $0$ and $1$, but this does not change any of the result qualitatively.}. We use a lognormal distribution, instead of a normal distribution, so that it is bounded below by $0$, and can, in principle, represent a physical density field. Note that we finally use the overdensity field $\delta_{\rm noise}$ in our calculations, which is bounded below by $-1$. $f_{\rm noise}$ is a parameter that lies between $0$ and $1$, and is used to control the relative noise level. For $f_{\rm noise} \sim 1$, the field $\tilde \delta$ is completely noise-dominated, while for small values of $f_{\rm noise}$, the cosmological signal dominates. For tracers, we once again choose the  $10^5$ most massive halos in the simulation volume. We study the cross-correlations between these halos and the $\tilde \delta$ field as a function of $f_{\rm noise}$, contrasting the cross-correlation detection through two-point measurements and through the nearest neighbor measurements from Sec.~\ref{sec:calculations}. Concretely, the quantity $f_{\rm noise}$ is varied between $[0.25,0.95]$ in steps of $0.1$. To test the significance of detection of the cross-correlation signal, we repeat the cross-correlation measurements, both two-point and nearest neighbor methods, using the original set of halos, and the matter overdensity field from $100$ \textit{different} realizations at the same cosmology, using the same values of $f_{\rm noise}$ to add noise to these matter fields. Since there should be no physical cross-correlation in the case of independent realizations, this exercise allows us to characterize the ``measurement'' noise for the uncorrelated case. To compute cross-correlations using the nearest neighbor method, we evaluate $\mathcal P_{\geq 1, >\delta_r^*}$ with the density threshold, $\delta_r^*$ set to the $50^{\rm th}$ percentile of the density values smoothed at each $r$.

The results are plotted in Fig.~\ref{fig:noise}. Since $f_{\rm noise}$ is varied over a large range, we plot the difference between the measurements in the correlated case (when the halos and the matter field are from the same realization) and the mean of the measurements in the $100$ uncorrelated samples (when the halos and the matter field come from different realizations), divided by the $1\sigma$ scatter in the measurement from the $100$ uncorrelated samples. This way of plotting enables a direct comparison of the relative signal-to-noise across different values of $f_{\rm noise}$. The left hand panel of Fig.~\ref{fig:noise} shows the results for the two-point cross-correlation measurements. The different shaded lines represent different values of $f_{\rm noise}$ - the darker lines correspond to higher values of $f_{\rm noise}$. The grey shaded region represents $1\sigma$ scatter in the uncorrelated measurements. Given the quantity that is being plotted, this scatter is always represented by a horizontal band between $-1$ and $1$, across different values of $f_{\rm noise}$. The inset zooms in on the $f_{\rm noise}=0.95$ measurement since its signal-to-noise is quite small. As expected, the signal-to-noise falls with $f_{\rm noise}$, across scales. From the inset, it is clear that when the noise fraction is very high, there is essentially no cross-correlation signal when using the two-point measurements. Even though it is not plotted in the inset, it should be noted that for $f_{\rm noise}=0.85$ as well, the signal in the two-point cross-correlation is quite small.

On the right panel of Fig.~\ref{fig:noise}, we plot the same quantities as above, but for cross-correlations evaluated using the nearest neighbor measurements. Similar to the left panel, the signal-to-noise of the nearest neighbor cross-correlation goes down with increasing amounts of noise added to the matter field. However, as shown in the inset of the right panel, even when $f_{\rm noise}=0.95$, i.e. the relative weight of the noise is $20$ times that of the true signal, the cross-correlation signal from the nearest neighbor measurements (green line) is clearly separated from the uncorrelated cases (grey band). Note that for $f_{\rm noise}=0.8$, the signal is  evident in the main panel itself --- starting at $\sim 40$ on the small scales. 

To quantitatively contrast the signal-to-noise from two-point cross-correlations and nearest-neighbor cross-correlations in the $f_{\rm noise}=0.95$ case, we compute the $\chi^2$ value from each statistic:
\eq{chi_sq}{\chi^2 = \Big(\psi^{\rm corr}-\bar \psi^{\rm uncorr}\Big)^T\mathbf C^{-1}\Big(\psi^{\rm corr}-\bar \psi^{\rm uncorr}\Big) \, ,}
where $\psi$ is the data vector of the statistic under consideration --- two-point cross-correlation or NN cross-correlation, evaluated with the appropriate noise level added to the matter field. $\psi^{\rm corr}$ is the data vector for the correlated case, while $\psi^{\rm uncorr}$ is the mean of the data vector from the $100$ uncorrelated realizations. The covariance matrix $\mathbf C$ is constructed from the $100$ uncorrelated realizations. We use an appropriate Hartlap factor \citep{2007A&A...464..399H} to account for the fact that we are estimating the covariance matrix from a finite number of realizations. Note that, unlike the visual representation in Fig.~\ref{fig:noise}, the covariance matrix takes into account the correlations in measurements between different radial bins. In this case, the bins are spaced equally between $10\hmpc$ and $30\hmpc$. The nearest neighbor measurements typically have more correlations between different bins than the two-point cross-correlation, which will be reflected in significant off-diagonal terms in the covariance matrix. The final $\chi^2$ value, therefore, better captures the total signal-to-noise than the purely diagonal information in Fig.~\ref{fig:noise}. We find that the $\chi^2$ value from the 2-point cross-correlation is $20.58$, while that from the nearest-neighbor cross-correlation is $127.97$. Measurements from $20$ different radial bins were used in each case, so the formal number of degrees of freedom is $19$. Assuming that that likelihood is Gaussian for both sets of statistics, we convert the obtained $\chi^2$ values into a $p$-value, which captures the probability of getting a $\chi^2$ value greater than that obtained above, but in the case when the halo sample and the matter field are uncorrelated. For the two point cross-correlation, the $p$-value is $0.36$. This implies that the cross-correlation is not detected at any statistically significant level. On the other hand, for the nearest neighbor cross-correlation, the $p$-value is $<10^{-7}$, and so the cross-correlation signal is detected at $>5\sigma$, or highly significant  level. It is important to note that the scales used in the analysis lie in the quasi-linear regime, rather than the fully nonlinear regime. The detection of a signal on the quasi-linear scales can typically be modeled with fewer theoretical systematics. In Sec.~\ref{sec:HEFT}, we show that the higher range of scales ($\geq 20 \hmpc$) used in this calculation, can already be faithfully modeled with existing methods.

This example demonstrates at the same levels of noise in the continuous field, the nearest-neighbor method of computing halo-matter cross-correlations yields higher signal-to-noise than that from two-point cross-correlation measurements. This is not entirely surprising, since the former is sensitive to all the higher order cross-correlations between the spatial fluctuations of the halo positions and the true underlying signal that are generated through gravitational evolution. Since the noise field is entirely uncorrelated with the underlying signal, all higher order cross-correlations must come from the true matter field, and therefore the nearest-neighbor method is much more efficient at isolating the signal, than the two-point cross-correlation which captures the lowest order. It is also worth noting that we have only used the $k=1$ nearest neighbor measurements to compute the cross-correlations --- in principle, there are also cross-correlation signals for higher $k$ nearest neighbor measurements, which can further enhance the total signal-to-noise.

\section{Modeling Tracer-Field cross-correlations with Hybrid Effective Field Theory}
\label{sec:HEFT}

In this section, we turn to the theoretical modeling of the cross-correlation measurements between tracers, such as halos or galaxies, and continuous fields, such as the total matter field, obtained from the formalism presented in Sec.~\ref{sec:formalism}. Since the large (linear) scales remain Gaussian, the gains from the $k$NN cross-correlations over the two-point cross-correlations are expected to be significant only at the quasi-linear to fully nonlinear scales, and so modeling efforts should be focused on these scales. We work in the framework of Hybrid Effective Field Theory, introduced in \cite{modichenwhite}. We summarize the main features of HEFT in Sec.~\ref{sec:HEFT_model}, and in Sec.~\ref{sec:HEFT_calculations}, we outline the procedure for specifically modeling tracer-field $k$NN cross-correlations in HEFT framework.

\subsection{Biased tracers and Hybrid Effective Field Theory}
\label{sec:HEFT_model}

 To do cosmology with a set of arbitrarily chosen tracers of the underlying matter distribution of the Universe, we require a prescription for the \textit{tracer-matter} connection, which allows for the correct marginalization over unknown properties of the chosen tracer sample, and multiple such approaches exist in the literature \cite[see][ for a recent review of the field]{wechslertinker}. Here, we will focus on an approach appropriate for modeling the tracer-matter connection on quasi-linear scales --- the bias expansion approach \cite[see][for a review of the subject]{dejacquesetal}. In particular, we will use a second-order Lagrangian bias expansion, with displacements computed from $N$-body simulations, as opposed to perturbation theory. \cite{modichenwhite} demonstrated that this hybrid approach, which is referred to as Hybrid Effective Field Theory (HEFT) in the literature, allows for smaller scales (down to $\sim 10\hmpc$ for two point correlation functions) to be modeled correctly at the same order in the bias expansion. The method has since been further developed to account for the joint variation of the bias and cosmological parameters \citep{Kokron2021,2021arXiv210112187Z} in real and redshift \citep{2022arXiv220706437P} space, and applied to the analysis of Dark Energy Survey (DES) Y1 data \citep{boryana}. While these works focused on two-point statistics, both auto and cross-correlations, \cite{kNN_HEFT} demonstrated that the same Formalism, with the same of parameters, can be used to model the $k$-nearest neighbor distributions of tracers. Crucially, both the two-point functions and the nearest neighbor distributions were fit the \textit{same values} of the model parameters, providing a consistency check for the model, as well as proving that these different summary statistics could be jointly modeled in a full analysis.

Here, we will demonstrate that the HEFT formalism can also be used to model the cross-correlation measurements between discrete tracers (halos) and continuous fields (the total matter field in the simulation)\footnote{\cite{kNN_HEFT} also considered the joint $k$NN distribution, or the cross-correlation between tracers of simulation particles as tracers of the matter field, and halos, but the entire formalism was in terms of discrete tracers.}. In the model, we assume that in Lagrangian (or initial) coordinates $\bq$, the tracer overdensity is related to various operators of the dark matter field that are allowed by the symmetries of the system \citep{Vlah_2015}. In particular, at second order, 
\begin{align}
    \label{eqn:lagbias}
    \delta_X (\bq) &= F[\delta(\bq), s_{ij} (\bq)] \\
    &\approx 1 + b_1 \delta(\bq) + b_2 \left (\delta^2 (\bq) - \langle \delta^2 \rangle) \right ) + \nonumber  \\
    & \quad\quad b_{s^2} \left (s^2 (\bq) - \langle s^2 \rangle) \right ) + b_{\nabla^2} \nabla^2 \delta (\bq) + \cdots  \nonumber\\
    & \quad+ \epsilon (\bq). \nonumber
\end{align}
$\delta_X$ represents the overdensity field of tracer $X$. $s_{ij} (\bq) = \left ( \frac{\partial_i \partial_j}{\partial ^2} - \frac{1}{3}\delta_{ij} \right ) \delta (\bq)$ is the traceless part of the tidal tensor field, and $\epsilon (\bq)$ is the stochastic term, which is uncorrelated (at the 2-point level) with the other operators appearing in Eq.~\ref{eqn:lagbias}. The $b_i$ are the (scale-independent) bias parameters, and take on different values for different tracer samples. At low redshift, or late times, the tracer density in Eulerian coordinates is given by \citep{Matsubara_2008}
\begin{align}
    1 + \delta_X (\bx) = \int d^3 q F[\delta(\bq), s_{ij}(\bq)] \delta^D \left ( \bx - \bq - \bPsi(\bq) \right ) \, ,
	\label{eq:continuous_tracer_field}
\end{align}
with $\bPsi(\bq)$ representing the displacement vector connecting the initial Lagrangian coordinate $\bq$ to a final Eulerian coordinate $\mbx$. In the HEFT framework, the displacement vector $\bPsi(\bq)$ is taken directly from $N$-body simulations, and therefore contains information about the full nonlinear dark matter dynamics \cite[see e.g.][for implementation details]{Kokron2021,kNN_HEFT}. In essence, the HEFT formalism produces each of the operator fields in Eq.~\ref{eqn:lagbias} advected to late times. Note that since we have the final particle positions from the simulations, $\delta_m(\mbx)$, the late time total matter density field can also be easily computed. Given the late time operator fields, the clustering measurements on the set of tracers of interest can be matched by changing the values of the bias parameters. Note that since the advection of the operator fields themselves is independent of the actual values of the bias parameters $\{b_i\}$, evaluating the model for different $\{b_i\}$ is computationally inexpensive, and only needs to be done in post-processing of the simulation data.

\begin{figure*}
	\includegraphics[width=0.9\textwidth]{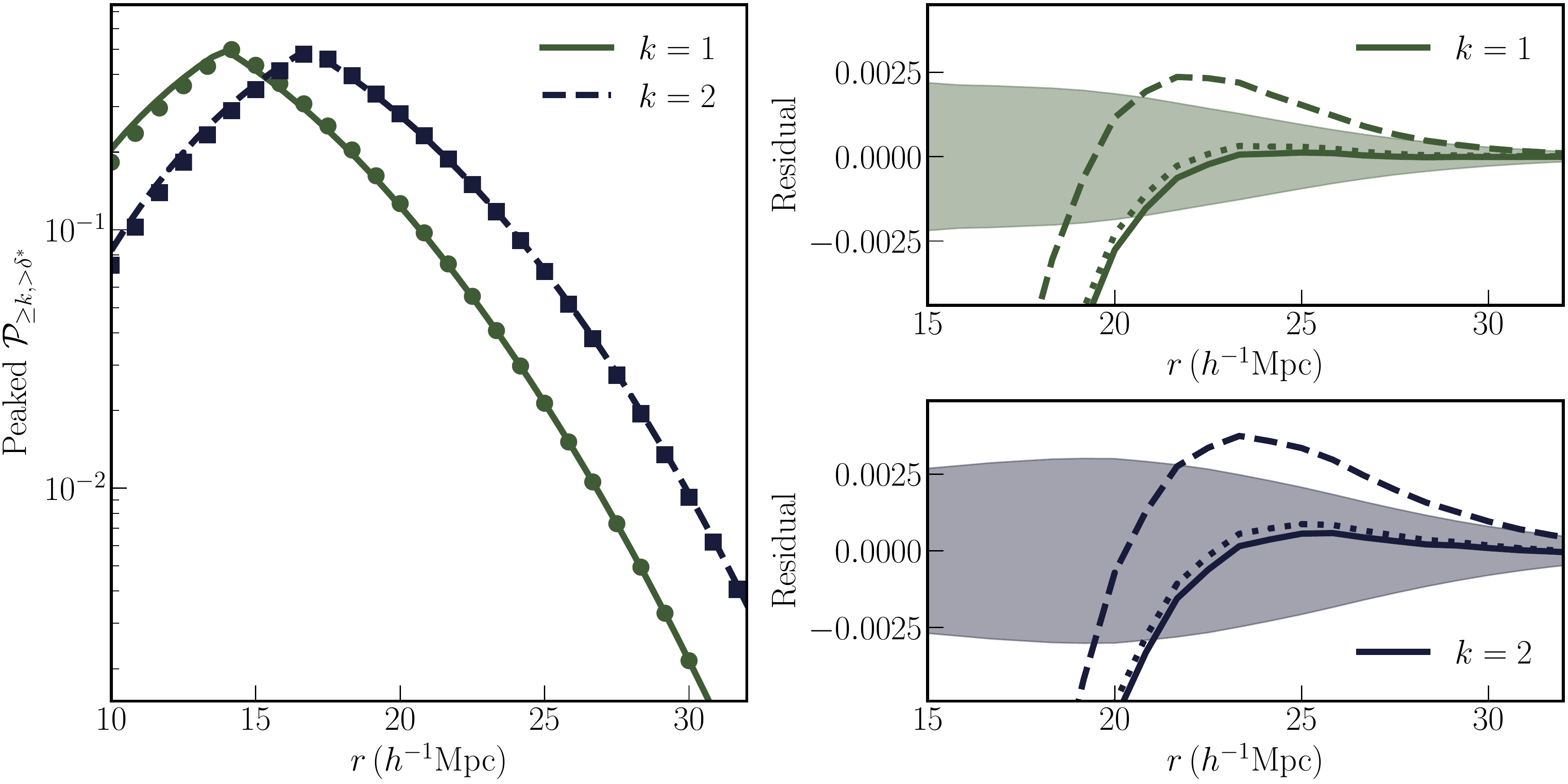}
	\caption{\textit{Left panel:}  Peaked version (see text for details) of the measurement of $\mathcal P_{\geq k,>\delta_r^*}$ with the solid green line for $k=1$, and the dashed blue line for $k=2$, for a set of $10^5$ halos from a $(1\hgpc)^3$ simulation, and the underlying matter field. $\delta_r^*$ is chosen using the constant mass method (see Sec.\ref{sec:calculations} for details) with $r_L=10\hmpc$. The solid green circles and the blue squares represent the predictions for the same measurements from the HEFT calculations. The values of the bias parameters in the HEFT calculations have been tuned to give the best fit to the measurements. Both the measurements and the HEFT predicitons have been averaged over 8 realizations at the same cosmology. \textit{Right panel}: Residual between the HEFT predictions and the actual measurements from the simulations. The shaded region represents the scatter in the measurements across 100 realizations at the same cosmology. $k=1$ is plotted on the top, and $k=2$ on the bottom. The solid line in each panel represents the residual when the bias parameters are tuned to provide the best fit to the $k$NN cross-correlation measurements, i.e. those corresponding to the curves in the left panel. The dashed lines represent the residual when the bias parameters are set to the values that provide the best fit to two-point measurements. The dotted lines represent the residuals when the bias parameters are set to the best fit of the $k$NN measurements of the halos alone. Note that the bias values from the solid lines remain a very good fit to the two-point measurements, even though the bias values from the two point measurements are not necessarily a good fit to the $k$NN measurements (dashed lines). The HEFT formalism is, therefore, able to model the $k$NN cross-correlations between halos and the matter field down to $\sim 20 \hmpc$.}
	\label{fig:HEFT}
\end{figure*}

\subsection{Computing Nearest Neighbor Tracer-Field Cross-Correlations in HEFT}
\label{sec:HEFT_calculations}

We choose a sample of $10^5$ halos from one of the \quijote boxes at the fiducial cosmology, following the process laid out in detail in \cite{kNN_HEFT}. The final catalog contains halos in the mass range $(10^{13.26}-10^{13.5}M_\odot/h)$, whose clustering is well described within HEFT \cite[see][]{modichenwhite}. We then compute the nearest-neighbor cross-correlations, using the $k=1$ and $k=2$ nearest neighbor measurements, between this halo sample and the matter density field, following the steps laid out in Sec.~\ref{sec:calculations}, and using query points on a $256^3$ grid, and the density field defined on a $1024^3$ grid. In this case, we use a ``constant mass" overdensity threshold for the matter density field, with the mass $M_0$ defined as the average mass enclosed in spheres of radius $10\hmpc$ in the simulation volume. We repeat the same measurements over $8$ realizations at the same cosmology, and average over them. The results are shown using the solid green line for $k=1$ NN and the solid blue line for $k=2$ NN on the left panel of Fig.~\ref{fig:HEFT}. Since the measurements asymptote to $1$ on the larger scales displayed in the plot, we plot the \textit{peaked} version of the values. This is defined in the following way: the value of $\mathcal P_{\geq k, >\delta_r^*}$ is plotted when  $P_{\geq k, >\delta_r^*}\leq 0.5$, and $1-\mathcal P_{\geq k, >\delta_r^*}$ is plotted when $\mathcal P_{\geq k, >\delta_r^*} > 0.5$. This way of plotting makes it easier to visually compare values on the right hand side of the plot, i.e. the behavior of $\mathcal P_{\geq k, >\delta_r^*}$ on larger scales.

Next, we outline the steps to make predictions for these measurements using the HEFT model. Given a set of values for the bias parameters $\{b_i\}$, we define the late time advected tracer field $\delta_X$, and the matter field from the simulation particles on a $1024^3$ grid. As described in detail in \cite{kNN_HEFT}, for the tracers alone, the values of the first $2$ nearest neighbor CDFs at radius $r$ are given by 
\eq{1nn_expression_matter}{\cdf_{1\nn} (r) &= \mathcal P_{\geq 1} (r) = \Big \langle 1 -  \exp \left[-\bar \lambda\left(1+\delta_{X,r}\right ) \right] \Big\rangle \\ \cdf_{2\nn} (r)&= \mathcal P_{\geq 2}(r) = \Big \langle 1 -  \exp \left[-\bar \lambda\left(1+\delta_{X,r}\right ) \right] \nonumber \\ & \qquad -   \bar \lambda \left( 1+ \delta_{X,r}\right) \exp \left[-\bar \lambda\left(1+\delta_{X,r}\right ) \right] \Big\rangle \label{eq:2nn_expression_matter}\, ,}
where $\delta_{X,r}$ represents the value of $\delta_X$ smoothed on radius $r$ at each point on the grid. The average, represented by $\langle ... \rangle$, is taken over all grid points. $\bar \lambda = 4/3 \pi r^3 \bar n$ is set by the mean number density of the tracers. Note that the two equations above are simply capturing the fact that the tracers number density represents a local Poisson process on the field $\delta_{X,r}$. Changing the values of $\{b_i\}$ yield different configurations of $\delta_X$, and thereby change the predictions for the nearest neighbor CDFs. To compute the cross-correlation of tracers drawn from $\delta_X$ with the matter field $\delta$, we are interested in the quantities like $\mathcal P_{\geq 1,>\delta_r^*}(r)$ and $\mathcal P_{\geq 2,>\delta_r^*}(r)$, the expressions need to be modified to read
\eq{updated_expression}{\mathcal P_{\geq 1, >\delta_r^*} (r) &=  \Big \langle \Big( 1 -  \exp \left[-\bar \lambda\left(1+\delta_{X,r}\right ) \right] \Big ) \Theta \big(\delta_r - \delta_r^*\big) \Big\rangle \\
\mathcal P_{\geq 2, >\delta_r^*} (r) &=  \Big \langle \Big( 1 -  \exp \left[-\bar \lambda\left(1+\delta_{X,r}\right ) \right] \nonumber \\ 
 & -   \bar \lambda \left( 1+ \delta_{X,r}\right) \exp \left[-\bar \lambda\left(1+\delta_{X,r}\right ) \right] \Big ) \Theta \big(\delta_r - \delta_r^*\big) \Big\rangle \, ,}
 where $\Theta(\delta_r - \delta_r^*)$ is the Heaviside step function, which accounts for the thresholding on $\delta_r$. Once again, the subscript $r$ represents the matter density $\delta$ and the HEFT tracer field $\delta_X$ smoothed on scale $r$, using a spherical top hat function. This smoothing is done in Fourier space (using an FFT approach), using the same $1024^3$ grid on which the overdensities are originally defined. 
 
 The results for the best-fit values\footnote{The best-fit bias values, for this first study, are determined by inspection, rather than a rigorous fitting procedure.} of $\{b_i\}$ are plotted in the left panel of Fig.~\ref{fig:HEFT} using the green circles for $k=1$NN and the blue squares for $k=2$NN. To facilitate easier comparison, we plot the residual between the measurements and the HEFT predictions on the right panels --- $k=1$ on the top panel and $k=2$ on the bottom panel. The solid curves in each panel correspond to the residual for the case where the best-fit values of $\{b_i\}$ obtained by fitting the $k$NN cross-correlation measurements, i.e. those corresponding to the left panel. The dashed curves on the right hand panels represent the residuals for the case where values for $\{b_i\}$ that go into generating the HEFT $k$NN cross-correlation predictions are obtained by fitting the two-point auto and cross-power spectra for the same set of halos, i.e. using the setup presented in \cite{Kokron2021}. Finally, the dotted curves represent the residuals between the measurements and predictions when the $\{b_i\}$ used to generate the HEFT cross-correlation predictions are obtained from fitting the $k$NN-CDFs of the tracers \textit{only}, i.e. the auto clustering alone. This is the same as the best-fit values for the identical tracer sample presented in \cite{kNN_HEFT}. The shaded regions on the right panels of the figure represent the realization-to-realization scatter in the $k$NN tracer-field cross-correlation measurements, estimated from $100$ realizations at the fiducial \textsc{Quijote} cosmology.

 From the solid lines on the right panel of Fig.~\ref{fig:HEFT}, we conclude that the HEFT framework is useful in modeling the halo-matter $k$NN cross-correlations down to $\gtrsim 20 \hmpc$, breaking down at slightly larger scale for $k=2$ than $k=1$. The comparison between the solid and the dotted lines on the right panel of Fig.~\ref{fig:HEFT} show that the best-fit HEFT model from the nearest neighbor measurements of the tracers \textit{alone} is completely consistent with the best-fit bias parameters obtained from the tracer and matter field NN cross-correlations. Since the latter is formally sensitive to the full joint distribution of the tracers and the matter field, this agreement serves as a powerful diagnostic of the reach of the HEFT formalism, as well as its physical underpinnings, with the halo field written in terms of operators of the dark matter field. The fact that the HEFT predictions from the best-fit $\{b_i\}$ to the 2-point auto and cross-correlations do not agree with the $k$NN cross-correlation measurements, as evidenced by the dashed curves on the right panel falling outside the shaded regions, illustrates the greater sensitivity of the $k$NN measurements to the underlying bias parameters. To go from the dashed lines on the figure to the solid lines, we only had to change the value of $b_{s^2}$ by about $12\%$ of its value from the 2-point best-fit. All other bias parameters were left unchanged. This is fully consistent with what was found in \cite{kNN_HEFT} for the $k$NN distributions of the tracers only. It should be noted that ths change in the value of $b_{s^2}$ changes the HEFT predictions for the 2-point functions only marginally, and the updated predictions remain extremely good fits to the 2-point function measurements given the measurement error bars.

 The HEFT framework, at second order in Lagrangian bias, therefore, allows for the consistent modeling of  nearest-neighbor halo-matter cross-correlations, the nearest-neighbor distributions of the halos alone, and of the two-point auto and cross-correlations on quasi-linear scales. By combining these measurements, it is possible to constrain the bias parameters much better than from the two-point measurements alone. This, in turn, allows for much better constraints on cosmological parameters of interest. Given the potential gains on parameter constraints, therefore, it is imperative to conduct modeling efforts in the presence of various survey systematics. This will be taken up in future work.

\section{Summary and Discussion}
\label{sec:conclusions}

In this paper, we have developed the formalism for describing correlated spatial fluctuations between the number counts of a set of tracers and the value of a continuous field by combining $k$-nearest neighbor distances \citep{Banerjee_Abel} for the tracers with measurements of the continuous field smoothed on various scales. The fundamental quantity that is computed is $\mathcal P_{\geq k, \delta_r^*}$, the volume-averaged probability of finding more than $k$ tracers within radius $r$ of a point \textit {and} the value of the density field exceeding a threshold $\delta_r^*$ when smoothed on scale $r$ around the point. Unlike the two-point tracer-field cross-correlations that are used ubiquitously in various cosmological applications, the cross-correlation method developed here is formally sensitive to all higher order (beyond linear or Gaussian) correlations between the two datasets. In fact, this particular cross-correlation only vanishes when the fluctuations in the two datasets are statistically independent - a much more stringent condition that a vanishing two-point cross-correlation. After developing the formalism and measurement procedure in Sec.~\ref{sec:formalism}, we have demonstrated the sensitivity to higher order, i.e. beyond Gaussian, cross-correlations with a concrete example in Sec.~\ref{sec:examples}. We have also demonstrated that for a fixed level of noise contaminating the continuous field under consideration, the nearest-neighbor cross-correlations of a set of tracers (halos) with the field has much higher signal-to-noise than the two-point cross-correlation. This represents an important result, with potential application to stronger detection of, e.g., the cross-correlation of the cosmological $21$cm emission signal with discrete tracers of LSS, such as galaxies and quasar samples. We will study this in greater detail in future work. Finally, in Sec.~\ref{sec:HEFT}, we have explored how the nearest-neighbor cross-correlation measurement of halos and the underlying cosmological matter density field can be modeled on quasi-linear scales with the Hybrid Effective Field Theory setup \citep{modichenwhite,2021arXiv210112187Z,Kokron2021}. Within this setup, we have shown that for scales $\geq 20 \hmpc$, the measurements can be modeled with the same set of bias parameters that are used to model the two-point functions, and have been shown also to model the $k$NN-CDFs of the tracers alone in \cite{kNN_HEFT}. In fact, the best-fit values of the individual bias parameters from fitting the $k$NN-CDFs of the tracers alone are completely consistent with the best-fit values of the same parameters using the cross-correlation measurement developed in this paper. Therefore, the auto and cross clustering of halos with the matter field can be consistently modeled within the HEFT framework. Given that no new parameters are needed in the modeling, and the additional statistical power of the nearest neighbor measurements over two-point measurements, constraints on various cosmological parameters may be considerably improved with the application of this method.

In most applications discussed here, the continuous field corresponds to the matter density field, or some physical density field, in general. This means that the threshold overdensity, $\delta_r^*$, used to evaluate $\mathcal P_{\geq k, >\delta_r^*}$ can be chosen to correspond to either a constant mass threshold, or a constant density percentile, or a constant overdensity value across scales, depending on the application. However, we stress that the continuous field need not necessarily represent a physical density field, as demonstrated in the example in Sec.~\ref{sec:beyond_Gaussian}. The only difference is this case is that ``enclosed mass" within radius $r$ is not well-defined, and therefore the constant mass thresholding is not useful. Instead the constant percentile or the constant value thresholds are more appropriate in such situations. It should also be noted that while we have used  3-dimensional fields and tracer positions in 3 dimensions, all our techniques generalize to 2 dimensions. Therefore, it is quite easy to use the nearest neighbor tracer-field cross-correlations on the projected matter field, e.g. weak lensing measurement, and  projected tracer counts. The modifications that are needed are to evaluate nearest neighbor distances in 2 dimensions instead of 3, and to use the 2-dimensional top-hat smoothing kernel on the field, instead of the 3-dimensional one.

The ``continuum limit'' of $k$NN-CDFs, as outlined in Sec.~\ref{sec:continuum_limit} can be used directly to measure the auto and cross-correlations of continuous fields. While there exist other techniques in the literature to analyze the auto-clustering of continuous fields which are also formally sensitive to clustering at all orders, such as the analysis of the full PDF \cite[e.g.][]{2020MNRAS.495.4006U} or Minkowski functionals \cite[e.g.][]{1989ApJ...340..647R,1998MNRAS.297..355S}, there are relatively few studies which do the same for cross-correlations of continuous fields \cite[see, however][]{2014MNRAS.442...69M}. The formalism presented here for cross-correlations between tracers and one continuous field generalizes easily to the case of two continuous fields, very similar to the statistics proposed in \cite{2014MNRAS.442...69M} for cross-correlating SZ and weak lensing maps. This aspect will explored in further detail in \cite{desknn} in the context of weak lensing measurements in the Dark Energy Survey (DES) across different redshift bins.

In Sec.~\ref{sec:HEFT}, we have demonstrated that the HEFT framework is able to model the halo-matter nearest-neighbor cross-correlations on intermediate scales. It is interesting to note that $\mathcal P_{\geq k, <\delta_r^*}$, which can be written in terms of the \textit{joint} PDF of the matter field and the halo field, from which the halo positions can be thought of as being generated through a local Poisson process, can formally be written in terms all $N$-point functions or cumulants of the halo field, all cumulants of the matter field, \textit{and} all joint cumulants of the two fields. Therefore, the auto-clustering information of the halos and the matter field are, in principle, present in the $\mathcal P_{\geq k, >\delta_r^*}$ measurements. By measuring at different $k$ and threshold $\delta_r^*$ it should be possible to constrain all the bias and cosmological parameters without having to separately analyze the $k$NN-CDFs of the halos alone, or the smoothed measurements on the matter field alone. This implies a reduction in the length of the data vector, essentially allowing for better compression of the overall data. Among other things, a reduced data vector length reduces the requirements on the number of realizations needed to compute the covariance matrix, assuming that the covariance matrix cannot be modeled analytically. This can be very helpful, since running the $N$-body simulations are the most computationally expensive step in setting up an analysis pipeline for such higher order methods. How well this data compression works in practice, and the potential gains, will be investigated in future work.

\section*{Acknowledgements}
The authors thank Dhayaa Anbajagane, Chihway Chang, Simon Foreman, Josh Frieman and Francisco Villaescusa-Navarro for helpful discussions and Dhayaa Anbajagane and Simon Foreman for helpful comments on an earlier version of this manuscript.
This work was partially supported by the U.S. Department of Energy SLAC
Contract No. DE-AC02-76SF00515. Some of the computing for this project was performed on the Sherlock cluster. The authors would like to thank Stanford University and the Stanford Research Computing Center for providing computational resources and support that contributed to these research results. The support and the resources provided by PARAM Brahma Facility under the National Supercomputing Mission, Government of India at the Indian Institute of Science Education and Research; Pune are gratefully acknowledged. The \textsc{Pylians3}\footnote{https://github.com/franciscovillaescusa/Pylians3} \citep{2018ascl.soft11008V} and \textsc{nbodykit} \citep{Hand_2018} analysis libraries  were used extensively in this paper. 

\section*{Data Availability}

The simulation data used in this paper is publicly available at \url{https://github.com/franciscovillaescusa/Quijote-simulations}. Additional data is available on reasonable request.



\bibliographystyle{mnras}
\bibliography{ref} 







\bsp	
\label{lastpage}
\end{document}